\documentclass[aps,reprint,showpacs, groupedaddress]{revtex4-1}

\usepackage{amsmath,amssymb,bm}
\usepackage{graphicx}
\usepackage[caption=false]{subfig}
\usepackage{bookman}
\usepackage{times}
\usepackage{ctable}
\usepackage{multirow}
\usepackage{floatrow}
\usepackage{wrapfig}
\usepackage{hyperref}

\begin{document}

\title[Non-Markovian electron dynamics in nanostructures]{Non-Markovian electron dynamics in nanostructures coupled to dissipative contacts}


\author{B. Novakovic}\email{novakovic@wisc.edu} 
\author{I. Knezevic}\email{knezevic@engr.wisc.edu}
\affiliation{Department of Electrical and Computer Engineering, University of Wisconsin -- Madison, Madison, WI 53706, USA}
\begin{abstract}
  In quasiballistic semiconductor nanostructures, carrier exchange between the active region and dissipative contacts is the mechanism that governs relaxation. In this paper, we present a theoretical treatment of transient quantum transport in quasiballistic semiconductor nanostructures, which is based on the
  open system theory and valid on timescales much longer than the characteristic relaxation time in the contacts.
  The approach relies on a model interaction between the current-limiting active region and the contacts, given in the scattering-state basis. We derive a non-Markovian master equation for the irreversible evolution of the active region's many-body statistical operator
  by coarse-graining the exact dynamical map over the contact relaxation time. In order to obtain the response quantities  of a nanostructure under bias,
  such as the potential and the charge and current densities, the non-Markovian master equation must be solved numerically together with the Schr\"{o}dinger, Poisson, and continuity equations.
  We discuss how to numerically solve this coupled system of equations and illustrate the approach on the example of a silicon \emph{nin} diode.
\end{abstract}
\maketitle                   






\section{Introduction}\label{sec:Introduction}

In nanoscale, quasiballistic electronic systems under bias, the process of
relaxation towards a nonequilibrium steady state cannot be
attributed to scattering, because these structures are small
compared to the carrier mean free path \cite{Lundstrom00,FerryGoodnick}. Rather, the active region of a nanostructure is an open quantum-mechanical system that exchanges
particles and information with the dissipative reservoirs of charge, usually
referred to as contacts \cite{Fischetti98,Fischetti99}. While qualitatively clear, a quantitative description of the irreversible evolution of the electronic system in this regime, where dissipation in the contacts coupled with the carrier exchange between the active region and contacts is the mechanism governing relaxation, is very challenging \cite{Ferry03,Ferrari04,Gebauer04}.

In this paper, we present a theoretical treatment of the transient-regime evolution of the  electronic system in a two-terminal ballistic nanostructure coupled to dissipative contacts and illustrate it on the example of a semiconductor \textit{nin} diode. The approach is rooted in the open system theory \cite{Alicki87,Breuer02}. We start from the closed-system, Hamiltonian dynamics of the many-body statistical operator for the ballistic active region and the dissipative contacts together, with a model interaction describing the injection of electrons into the active region. The model interaction Hamiltonian
differs from those typically employed \cite{Meir92,Jauho94}: it is specifically constructed to conserve
current during the process of carrier injection from/into the contacts, and its matrix elements are readily calculated from the
single-particle transmission problem for structures with and without resonances alike (Sec. \ref{sec:interaction}).
As is commonly done, we trace out the contact degrees of freedom and obtain the exact non-Markovian dynamical map that describes the evolution of the active region's statistical operator. However, while exact, this map is not useful in practical calculations. In order to obtain a tractable theoretical approach, we employ the fact that relaxation in the contacts of a nanostructure typically occurs on the shortest timescales in the whole system. We assume that the contacts are highly doped, so the fastest scattering mechanism is electron-electron scattering \cite{Lugli85,Osman87,Kriman90}. Within the momentum relaxation time,
the contacts adjust themselves to the new level of current flowing through the structure. The momentum relaxation time is virtually instantaneous from the standpoint of the nanostructure as a whole; if we are not to look into the microscopic details of relaxation in the contacts, but want to include their effect on the overall evolution of the nanostructure, the momentum relaxation time can be considered the shortest meaningfully resolvable time. Therefore, we coarse-grain the evolution over the contact momentum relaxation time and obtain a dynamical map that is piecewise Markovian but globally a non-Markovian, completely positive map (Sec. \ref{sec:master eqn}). We present a numerical algorithm for the calculation of relevant response quantities such as the charge density, potential, and current density based on the presented model, and illustrate the approach with a calculation of the response of a realistic semiconductor \textit{nin} diode in Sec. \ref{sec:transient}.

\section{Interaction between the active region and the contacts}\label{sec:interaction}

It has been well-established that the active region of a
nanostructure is an open quantum-mechanical system
\cite{Frensley90,Potz89}. Usually, the effect of openness is treated
through open boundary conditions; examples of such treatment are the
explicit source terms in the density matrix \cite{Brunetti89} or
Wigner function formalisms \cite{Frensley90,Nedjalkov04}.
Alternatively, a dynamical quantity is ascribed to the coupling
between the active region and the contact: in the popular
tight-binding variant of the nonequilibrium Green's function
formalism, pioneered by Datta \cite{Datta92_1}, the active
region-contact coupling is described through a special self-energy
term. In the Meir-Wingreen \cite{Meir92,Jauho94} approach and its
derivatives, one employs a coupling Hamiltonian between the contacts
and the active region, but no general recipe is available for the
derivation of the Hamiltonian matrix elements. Also, this approach
has so far been applied only when the active region supports a small
number of discrete states, so the model has little practical value
for structures with no resonances, such as an \textit{nin} diode, or
to account for the continuum states in structures with mixed
spectrum, such as a double-barrier tunneling structure (also known as the
resonant-tunneling diode). We present an alternative interaction Hamiltonian that does nor require that a
structure a priori possesses resonances, and whose matrix elements
are straightforwardly derived from the single particle transmission
problem.

\begin{figure}
\centering{
\includegraphics[width=3 in]{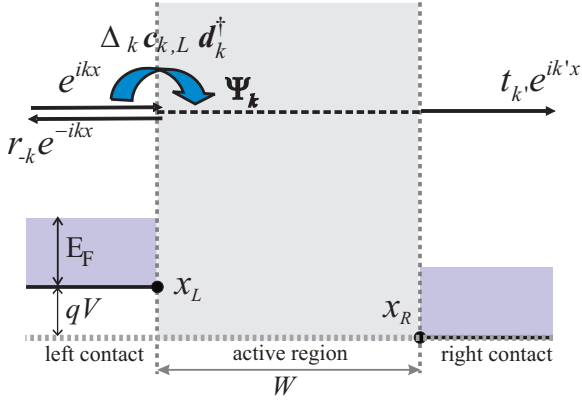}
\caption{ Schematic of the coupling between the active region of a
generic two-terminal nanostructure and the contacts. In the case of
ballistic injection through the open boundaries, a
forward-propagating state $\Psi_k$ is coupled with the state
$\exp(ikx)$ in the left contact
via a hopping model interaction (\ref{eq:Hplus}). After
\cite{Knezevic08}.}\label{fig:TwoTerminalSchematic}}
\end{figure}

Consider a generic two-terminal nanostructure under bias (Fig. 1).
For every energy $\mathcal{E}_k$ above the bottom of the left
contact, the active region's single particle Hamiltonian has two
eigenfunctions, a forward ($\Psi_k$) and a backward ($\Psi_{-k}$)
propagating state, that can be found by (in general numerically)
solving the single-particle Schr\"{o}dinger equation for a given
potential profile in the active region. Associated with $\Psi_k$ ($\Psi_{-k}$) in the active region are the
creation and destruction operators $d_k^\dagger$ and $d_k$
($d_{-k}^\dagger$ and $d_{-k}$), so the active region many-body
Hamiltonian is
\begin{equation}\label{eq:Hs}
\mathcal{H}_S=\sum_{k>0} \mathcal \omega_k (d_k^\dagger d_k
+d_{-k}^\dagger d_{-k}).
\end{equation}
Spin is disregarded, and $\omega_k=\mathcal{E}_k/\hbar$. In the case of
ballistic injection through the open boundaries, each state $\Psi_k$
is naturally coupled with the injected states $\exp(ikx)$ from the left
contact. For $\Psi_{-k}$, the coupling
is with $\exp(- ik'x)$ from the right contact ($k'^2-2meV/\hbar^2=k^2=2m\mathcal{E}_k/\hbar^2$, where $V$ is the applied bias).
To model this coupling via a hopping-type interaction,
we can write quite generally (see Fig.
\ref{fig:TwoTerminalSchematic})
\begin{eqnarray}\label{eq:Hplus}
\mathcal
H_{\mathrm{int}}=\sum_{k>0}\Delta_k d_k^{\dagger}c_{k,L}+\Delta_{-k} d_{-k}^{\dagger}c_{-k',L}+h.c.
\end{eqnarray}
$c^\dagger_{k,L}$ ($c_{k,L}$) and $c^\dagger_{-k',R}$
($c_{-k',R}$) create (destroy) an electron with a wavevector $k$
in the left and $-k'$ in the right contact, respectively. The hopping
coefficients $\Delta_k$ and $\Delta_{-k}$ are proportional to the current carried by each mode, i.e.
\begin{equation}\label{eq:Delta_k}
\Delta_k=\frac{I_k}{e\mathcal{T}_k},
\end{equation}
where $\mathcal{T}_k$ is the transmission coefficient of mode $k$. $\Delta_k$ can be written in terms of the
scattering-state injection amplitude \cite{Novakovic2012}.

\section{The transport master equation}\label{sec:master eqn}

In general, the dynamics of a nanostructure's active region is
non-unitary and non-Markovian (i.e., memory effects are important,
meaning that the system remembers how it got to a certain state and
its future direction of evolution depends not only on the state it
is currently in, but also on how it got to that state to begin
with). A non-Markovian, non-unitary map that would describe the
active region in a ballistic nanostructure in the presence of
contacts can be derived by tracing our the contact degrees of
freedom from the unitary evolution of the closed "active region and
contacts" system. A general form of the non-Markovian evolution of
the active region statistical operator $\rho$ is given by
$\rho(t)=\mathcal{W}(t,0)\rho(0)$, where map $\mathcal{W}$ is of the
form
\begin{equation}\label{eq:W in terms of K}
\mathcal{W}(t,0)=\mathrm{T^c}\exp{\left(\int_0^t
\mathcal{K}(t')\,dt'\right)}.
\end{equation}
Here, $\mathcal{K}(t)$ is the generator of the map
$\mathcal{W}(t,0)$. In general, it is impossible to obtain $\mathcal{W}(t,0)$ exactly.
If one is interested in retaining the non-Markovian nature of
(\ref{eq:W in terms of K}), typically an expansion up to the second
or fourth order in the interaction is undertaken \cite{Breuer02}. On
the other hand, a Markovian approximation to the exact dynamics can
quite generally be obtained in the weak-coupling limit. This limit has
been used previously by several authors \cite{Li05,Pedersen05} to
derive Markovian rate equations for tunneling structures in the
resonant-level model, although the weak-coupling approximation is not generally applicable
to nanostructures \cite{Li05}.

However, here we point out the Markovian approximation to the long-time evolution of
nanostructures can be justified more broadly, by employing
\textit{the approximation of a memoryless environment} for the
contacts. Basically, electron-electron
scattering in the highly doped contacts of semiconductor devices
ensures that the carrier distribution snaps into a drifted Fermi-Dirac distribution \cite{Lugli85}
within the energy-relaxation time $\tau\approx 10^1-10^2$
femtoseconds \cite{Osman87,Kriman90} (the actual value depends on
the doping density and temperature). This time is very short with
respect to the typical response times of these devices, which is on
the timescales of $\tau_{AR}\approx 1-10$ ps ("AR" stands for the
active region), so on these timescales contacts can be considered
\textit{memoryless}. For low-dimensional nanostructures, fabricated on a high-mobility
two-dimensional electron gas (2DEG) and operating at low
temperatures, phonons are frozen so the energy relaxation in the contacts is also governed
by the inelastic electron-electron scattering
\cite{Altshuler79,Altshuler81}. The ratio $\tau/\tau_{AR}$ is not as small
as in devices, but is still less than unity.

To practically obtain the Markovian approximation due to an
environment that loses memory after a time $\tau$, we use the
coarse-graining procedure: we can partition the time axis into
intervals of length $\tau$, $t_n=n\tau$, so the environment
interacts with the system in exactly the same way during each
interval $[t_n,t_{n+1}]$ \cite{Lidar01},
\begin{equation}\label{eq:difference rho_S}
\frac{d\rho_S}{dt}\approx\frac{\rho_{S,n+1}-\rho_{S,n}}{\tau}=
\mathcal{\overline{K}}_\tau\rho_{S,n},\end{equation} where
$\mathcal{\overline{K}}_\tau=\frac{\int_0^\tau
\mathcal{K}(t')dt'}{\tau}=\frac{\int_{t_n}^{t_{n+1}}\mathcal{K}(t')dt'}{\tau}$
is the averaged value of the map's generator over any interval
$[t_n,t_{n+1}]$ ($\mathcal{K}$ is reset at each $t_n$). If the
coarse-graining time $\tau$ is short enough, then the short-time
expansion of $\mathcal{K}$ can be used to perform the coarse-graining \cite{Knezevic08}, so we finally arrive at
the desired Markovian kinetic equation
\begin{eqnarray}\label{eq:Markov final}
\frac{d\rho_S(t)}{dt}=\left(-i\mathcal{L}_{\mathrm{eff}}-\Lambda\tau\right)\rho_S(t).
\end{eqnarray}
where ${\mathcal{L}}_{\mathrm{eff}}=[{\mathcal{H}}_S+\langle{{\mathcal{H}}_\mathrm{{int}}}\rangle,\dots]
=\mathcal L_S+[\langle{{\mathcal{H}}_\mathrm{{int}}}\rangle,\dots]$
is an effective system Liouvillian, containing the
noninteracting-system Liouvillian $\mathcal L_S$ and a correction
due to the interaction
[$\langle\dots\rangle=\mathrm{Tr}_E[\rho_E(0)\dots]$ denotes the
partial average with respect to the initial environmental state
$\rho_E(0)$]. The matrix elements of superoperator $\Lambda$, in a
basis $\alpha\beta$ in the system's Liouville space (Liouville space
is basically a tensor square of the Hilbert space), are determined
from the matrix elements of the interaction Hamiltonian:
\begin{eqnarray}\label{eq:Lambda}
{\Lambda}^{\alpha\beta}_{\alpha '\beta'}&=&\frac{1}{2}\left\{
\left\langle{{\mathcal{H}}}_{\mathrm{int}}^2\right\rangle^{\alpha}_{\alpha
'}\delta^{\beta '}_\beta +\left\langle
{{\mathcal{H}}}_{\mathrm{int}}^2\right\rangle^{\beta'}_{\beta
'}\delta^\alpha_{\alpha '}\right.\\
&-&2\sum_{j,j'}\left(
{{\mathcal{H}}}_{\mathrm{int}}\right)^{j'\alpha}_{j\alpha
'}\rho_E^j\left(
{{\mathcal{H}}}_{\mathrm{int}}\right)^{j\beta'}_{j'\beta}
-\left(\langle
{\mathcal{H}}_{\mathrm{int}}\rangle^2\right)^{\alpha}_{\alpha
'}\delta^{\beta'}_\beta\nonumber \\
&+&\left.2\langle
{\mathcal{H}}_{\mathrm{int}}\rangle^{\alpha}_{\alpha'}\langle
{\mathcal{H}}_{\mathrm{int}}\rangle^{\beta'}_\beta -\left(\langle
{\mathcal{H}}_{\mathrm{int}}\rangle^2\right)^{\beta'}_\beta\delta^{\alpha}_{\alpha
'}\right\},\nonumber
\end{eqnarray}
where $\rho_E^j$ are the eigenvalues of the initial environment
statistical operator $\rho_E(0)$. ${\Lambda}$ contains essential
information on the directions of coherence loss. Strictly speaking, the above coarse-graining procedure holds if
\begin{equation}\label{eq:validity of Markov}
\left\|\Lambda\right\| {\tau}^2 \ll
\min{\{1,\left\|\mathcal{L}_{\mathrm{eff}}\right\|\tau\}}.
\end{equation}

\noindent Since the interaction Hamiltonian is linear in the contact
creation and destruction operators, and we can approximate that each
contact snaps back to a "drifted" grand-canonical statistical
operator, we have $\langle \mathcal H_{\mathrm{int}}^{L/R}\rangle
=0$. This means that $\mathcal{L}_{\mathrm{eff}}={\mathcal{L}}_S$,
and also leaves us with only the first three terms in Eq.
(\ref{eq:Lambda}) for $\Lambda$ to calculate. It can be shown \cite{Knezevic08} that each term in $\Lambda$ is a sum of independent contributions over individual modes [$\Lambda=\sum_{k} \Lambda_k$] that attack only
single-particle states with a given $k$. The same holds for
$\mathcal {\mathcal{L}}_S$. Consequently, in reality we have a
multitude of two-level problems, one for each state $\Psi_{k}$,
where the two levels are a particle being in $\Psi_{k}$ ("+") and a
particle being absent from $\Psi_{k}$ ("-"). Each such 2-level
problem is cast on its own 4-dimensional Liouville space, with
$\rho_k=\left(\rho^{++}_k,\rho^{+-}_k,\rho^{-+}_k,\rho^{--}_k\right)^\mathrm{T}$
being the reduced statistical operator that describes the occupation of
$\Psi_{k}$. According to (\ref{eq:Markov final}),
\begin{equation}
\frac{d\rho_k}{dt}=[-i\mathcal
{\mathcal{L}}_{S,k}-\Lambda_k\tau]\rho_k,
\end{equation}
where
\begin{eqnarray}
{\mathcal{L}}_{S,k}=\left[\begin{array}{cccc}
0 & 0 & 0 & 0 \\
0 & 2\omega_k & 0 & 0 \\
0 & 0 & -2\omega_k & 0 \\
0 & 0 & 0 & 0 \end{array}\right],\quad {\Lambda}_k=\left[\begin{array}{cccc}
A_k & 0 & 0 & -B_k\\
0 & C_k & 0 & 0 \\
0 & 0 & C_k& 0 \\
-A_k & 0 & 0 & B_k\end{array},\right]
\end{eqnarray}
and
\begin{equation}
A_k=\Delta^2_{k}(1-f^{L}_k),\, B_k=\Delta^2_{k}f^{L}_k,\,C_k=(A_k+B_k)/2=\Delta^2_{k}/2 .
\end{equation}
The rows/columns are ordered as
$1=\left|+\right\rangle\left<+\right|,
2=\left|+\right\rangle\left<-\right|,
3=\left|-\right\rangle\left<+\right|,
4=\left|-\right\rangle\left<-\right|$. Clearly, off-diagonal elements $\rho_k^{+-}$ and $\rho_k^{-+}$
decay as $\exp{(\mp i2\omega_k-\tau C_k)t}$ and reach zero in the
steady state. The two equations for $\rho_{k}^{++}=f_k(t)$ and
$\rho_{k}^{--}=1-f_k(t)$ are actually one and the same, and either
one yields
\begin{subequations}\label{eq:f time evolution 0}
\begin{equation}
\frac{df_k}{dt}=-\tau(A_k+B_k)f_k+\tau B_k=-\tau\Delta_k^2 f_k +\tau \Delta_k^2 f_k^L,
\end{equation}
where $f_k$ is the distribution function for the active region. An analogous relationship holds for the backward-propagating states:
\begin{equation}
\frac{df_{-k}}{dt}=-\tau\Delta_{-k}^2 f_{-k} +\tau \Delta_{-k}^2 f_{-k'}^R
\end{equation}
\end{subequations}
Equations (\ref{eq:f time evolution 0}) may at first glance appear to be Markovian in form, but they are generally not, as we will discuss in the next section. However,
if we are in the low-bias regime and assume that: (1) the potential and thus the scattering states, transmission coefficients, and the coupling strengths $\Delta_{\pm k}$ are virtually constant throughout the transient, and (2) the current density is low, so any changes to the contact distribution functions that result from a current flow can also be neglected, then evolution (\ref{eq:f time evolution 0}) will indeed be Markovian \cite{Knezevic08}. In fact, we can solve the above equations analytically in the limit of low bias and low current densities. In that case, the contact distribution functions are nearly constant, and the solution to Eqs. (\ref{eq:f time evolution}) can be found as

\begin{eqnarray}\label{eq:low-field solution}
f_k(t)&=&\left(f_k(0)-f_k^L\right)e^{-\tau\Delta_k^2 t}+ f_k^L,\\ f_{-k}(t)&=&\left(f_{-k}(0)-f_{-k'}^R\right)e^{-\tau\Delta_{-k}^2 t}+ f_{-k'}^R.\nonumber
\end{eqnarray}

\noindent As expected, the steady-state values of the distribution functions are the contact distribution functions
\begin{equation}\label{eq:low-field solution}
f_k(\infty)=f_k^L,\quad f_{-k}(\infty)=f_{-k'}^R.
\end{equation}
A detailed discussion of the relationship of the model with the Landauer-B\"uttiker formalism can be found in \cite{Knezevic08}.

\section{Example: Transient in an \textit{nin} diode}\label{sec:transient}

As the current starts to flow through the structure, the contact distribution functions quickly adjust to accommodate the current flow. A good approximation for the distribution function in bulklike contacts in which electron-electron scattering is the most efficient mechanism
is the drifted Fermi-Dirac distribution function
\begin{equation}
f_k^L (k_d)=\frac{1}{\exp\left\{ \frac{\hbar^2[(k-k_d)^2-k_F^2]}{2m_{||}k_B T}\right\}+1},
\end{equation}
where $k_d$, the drift wave vector, depends on the total current density $J$ flowing through the structure as $k_d=m_{||}J/e\hbar n$. $m_{||}$ is the effective mass in the direction of current flow, and $n$ is the contact carrier density. $k_d$ changes during the transient and brings about non-Markovian character to Eqs. (\ref{eq:f time evolution}):
\begin{eqnarray}\label{eq:f time evolution}
\frac{df_k}{dt}&=&-\tau\Delta_k^2 f_k +\tau \Delta_k^2 f_k^L(k_d),\\
\frac{df_{-k}}{dt}&=&-\tau\Delta_{-k}^2 f_{-k} +\tau \Delta_{-k}^2 f_{-k'}^R (k_d).\nonumber
\end{eqnarray}
where it should be understood that $k_d$ changes with time.

As the transient progresses, the current and the charge density in the structure change, which in turn changes the potential profile, the scattering states available to electrons, the transmission coefficients,  and, to a small degree, the interaction matrix elements $\Delta_{\pm k}$, as well as the aforementioned contact distribution functions. Therefore, all these quantities have to be carefully updated during the simulation.

\begin{figure}[h]
\centering{\includegraphics[width=7 cm]{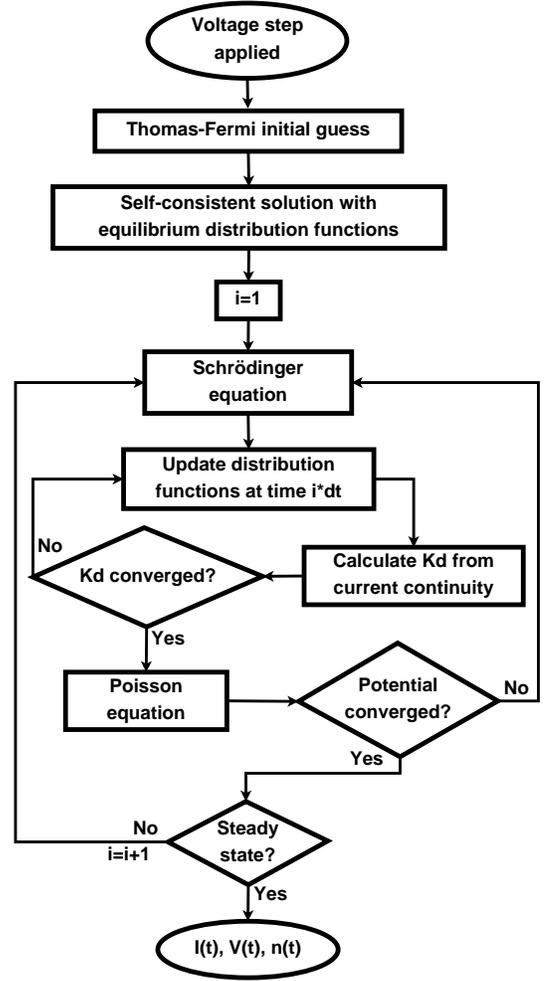}
\caption{Flowchart of the numerical algorithm for the calculation of the electronic response of a biased two-terminal nanostructure during a transient. }
\label{fig:flowchart}}
\end{figure}

Equations (\ref{eq:f time evolution}) must in general be solved numerically. A flowchart of the numerical algorithm is presented in Fig. \ref{fig:flowchart}. Upon the application of bias to the contact,
but before the current starts to flow ($t =0^{+}$), we solve the Schr\"{o}dinger and Poisson equations
with equilibrium initial distribution functions $f_{\pm k}(0)$.
At this point current continuity between the contacts and device is not necessary and $k_d=0$ ($J=0$).
We then proceed to the next time step, with non-zero current, and solve first the Schr\"{o}dinger equation
using the potential from the previous time step. Using the previous value for $k_d$, we update the distribution functions at the new time step and calculate the current and charge densities, then find the current density due to the change in the device charge density, iterate for a new $k_d$ until the current density in the contacts is equal to the sum of the current density in the device and the current density due to the change in the device charge density. In each new iteration, we use a $k_d$ that is formed as a weighted sum of $k_d$'s from the current and previous iterations. When the current and $k_d$ are self-consistently obtained, we use the newly obtained device charge density in the Poisson equation to obtain a new guess for the potential, and repeat until the potential converges (using the globally convergent Newton's method \cite{NumericalRecipesFortran} with a semiclassical Jacobian \cite{Lake97,Laux04}). We repeat for all time steps until a steady state is reached.

There are several nontrivial numerical considerations. One is the ability to achieve a high enough density of scattering states to properly represent physical quantities such as the charge and current densities or the potential profile. We are trying to capture a continuum of scattering states, which at first glance might seem doable by indiscriminately increasing the density of $k$'s by choosing larger and larger simulation domains. Unfortunately, this brute-numerical-force approach does not work; what does work instead is generating a "smart" discrete set of scattering states by first solving the Schr\"{o}dinger equation in the simulation domain with the condition that the first derivative be zero at the boundaries,
and then projecting these states onto the forward and backward moving scattering states. Details of this discretization of the scattering state continuum can be found in \cite{Laux04}.

\begin{figure}
\includegraphics[width=70 mm]{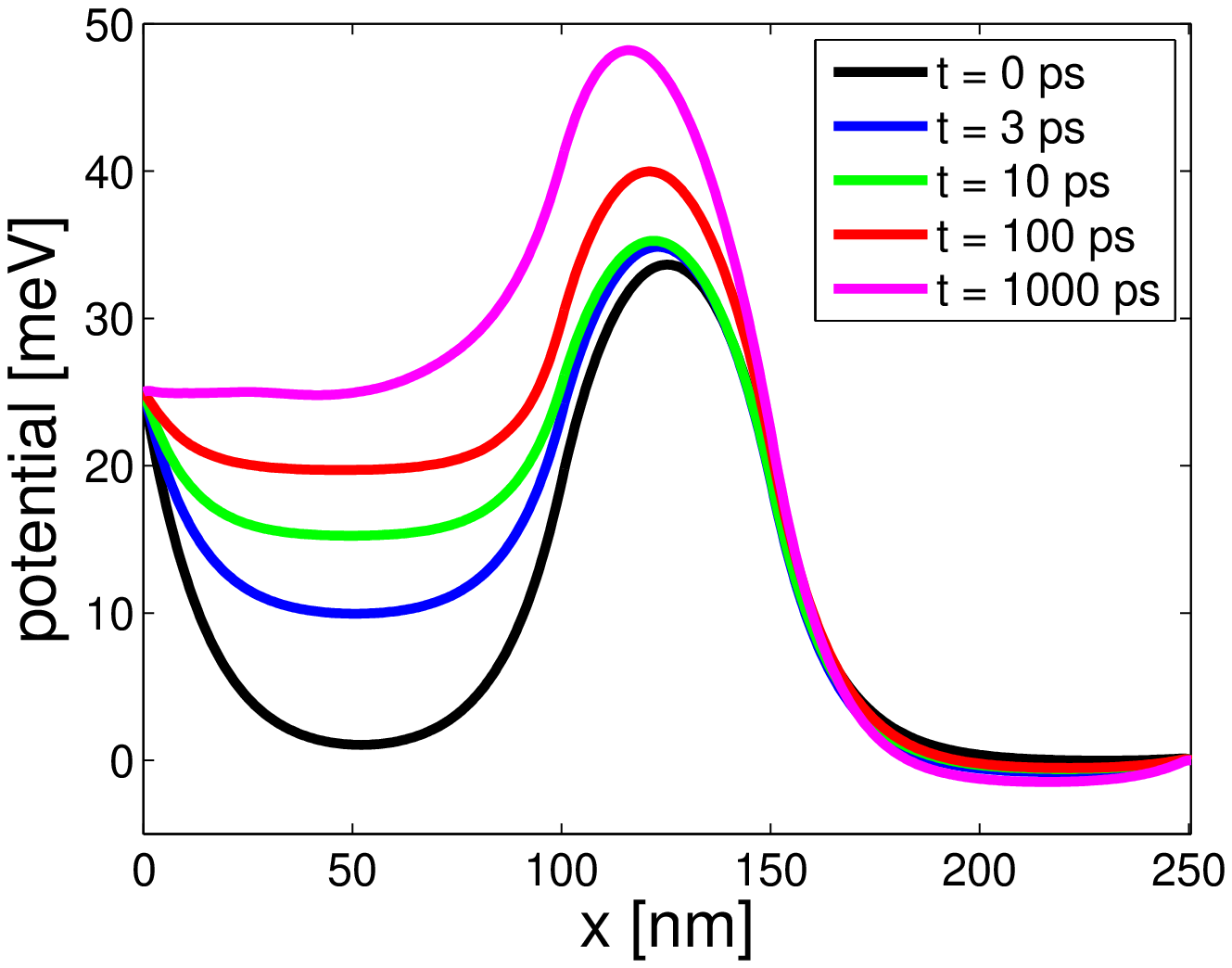}
\includegraphics[width=70 mm]{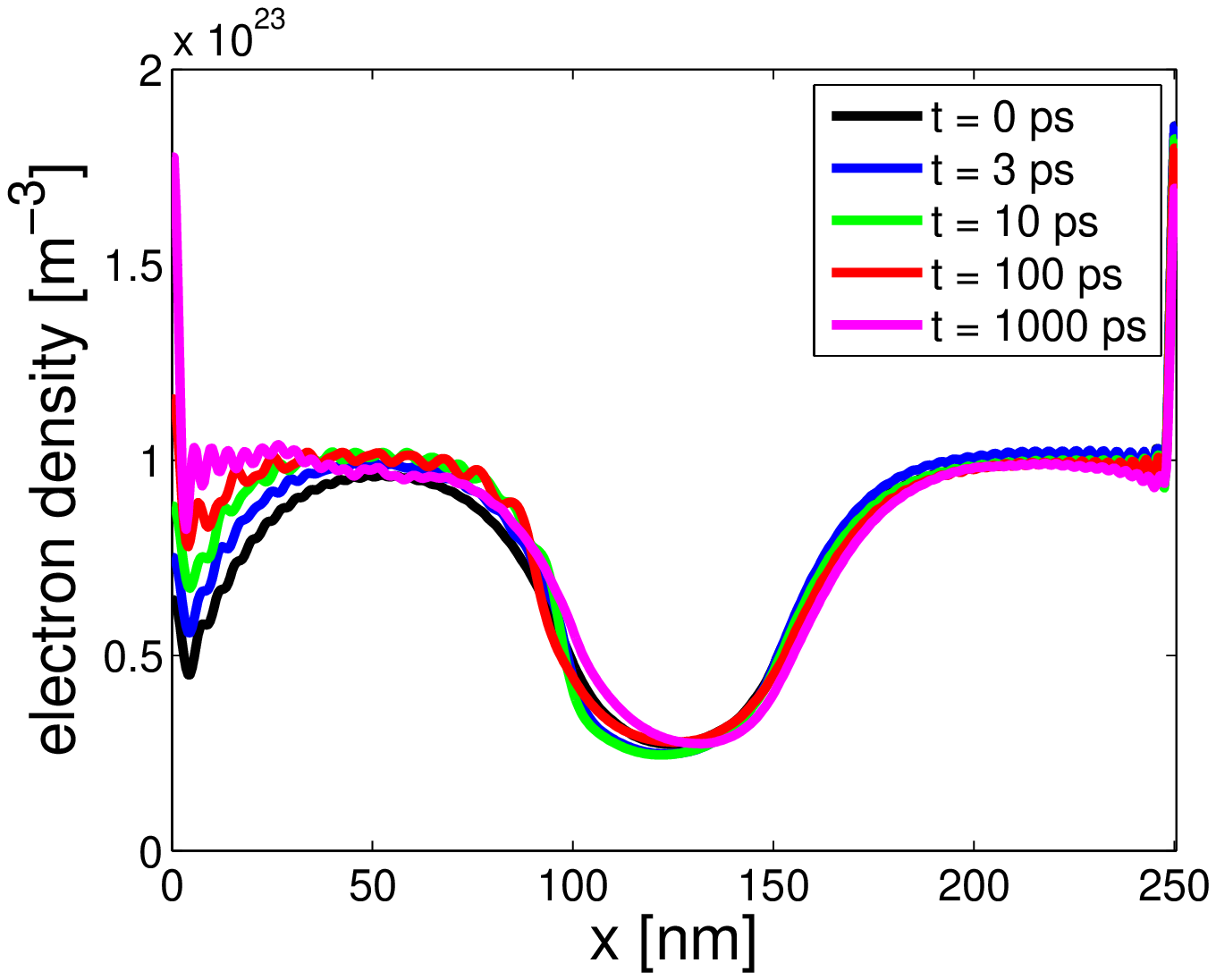}
\caption{(a) Potential and (b) charge density in the \textit{nin} diode as a function of time upon the application of -25 mV to the left contact. The \textit{n}-type regions are doped to $10^{17}$ cm$^{-3}$. The contact momentum relaxation time is $\tau$=120 fs, as calculated from the textbook mobility value corresponding to the contact doping density.}\label{fig:nin potential}
\end{figure}

A related question is how to populate a small set of bound states that can emerge in a biased nanostructure (e.g. note the potential pocket on the left-hand-side of Fig. \ref{fig:nin potential}a at times below 100 ps). Those states are in reality filled by electron-electron and electron-phonon scattering, essentially the same mechanisms as in the contacts. Since we are not treating scattering explicitly in this approach, we populate the bound states according to the Fermi level in the nearest contact. More detail on the finer points of the numerical simulation can be found in \cite{Novakovic2012}.

\begin{figure}
\centering{
\includegraphics[width=70 mm]{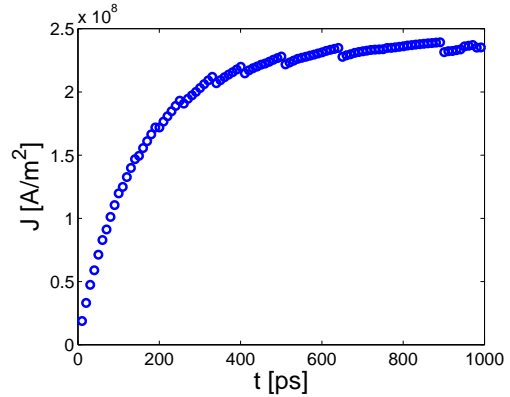}
\caption{Current density versus time for the \textit{nin} diode from Fig. \ref{fig:nin potential} upon the application of -25 mV to the left contact. The \textit{n}-type regions are doped to $10^{17}$ cm$^{-3}$ and  $\tau$=120 fs.}
\label{fig:nin current}}
\end{figure}

Figures \ref{fig:nin potential} and \ref{fig:nin current} depict the potential, charge density, and current density for a single ellipsoidal valley in an \textit{nin} silicon diode at room temperature. The left and right contacts are doped to $10^{17}$ cm$^{-3}$, whereas the middle region is intrinsic (undoped). The momentum relaxation time in the contacts is taken to be $\tau=$120 fs, based on the textbook mobility values for the
above doping density. Note that the characteristic response time of the current is of order hundreds of picoseconds, so three orders of magnitude greater than the contact relaxation time. The transient duration is long because of the relatively weak coupling between the active region and the contacts; the transient duration can be thought of as the inverse of a typical $\Delta_k^2\tau$ among the $k$'s participating in the current flow.

\section{Conclusion}\label{sec:Conclusion}

We presented a theoretical treatment of the transient-regime evolution of an
electronic system in a two-terminal ballistic nanostructure coupled to dissipative contacts.
The approach is rooted in the open system theory and is based on two key ingredients:
(1) A model interaction Hamiltonian between the active region and
the contacts, constructed specifically to conserve current during the process of carrier injection from/into the
contacts, whose matrix elements are readily calculated from the
single-particle transmission problem for structures with and
without resonances alike. (2) In the absence of scattering in the active region, it is the
rapid energy relaxation in the contacts (due to electron-phonon or, in good, highly-doped contacts, due to electron-electron scattering) that is the indirect source of irreversibility in the
evolution of the current-limiting active region, owing to the
contact-active region coupling. We account for the influence of the
rapid relaxation in the contacts by coarse graining the exact active region evolution over the contact momentum relaxation time.
The resulting equations of motion for the distribution functions of the forward and backward propagating states in the active region, Eqs. (\ref{eq:f time evolution}),
have non-Markovian character as they incorporate the time-varying contact
distribution functions through the time-dependent drift-wavevector that depends on the instantaneous current flowing. In order to obtain the response quantities  of a nanostructure under bias,   such as the potential and the charge and current densities, the non-Markovian master equations must be solved numerically together with the Schr\"{o}dinger, Poisson, and continuity equations. We presented an algorithm for the numerical solution of this coupled system of equations and illustrated the approach on the example of a silicon \emph{nin} diode.

\section{Acknowledgment}
This work has been supported by the NSF, award ECCS-0547415.

\providecommand{\WileyBibTextsc}{}
\let\textsc\WileyBibTextsc
\providecommand{\othercit}{}
\providecommand{\jr}[1]{#1}
\providecommand{\etal}{~et~al.}

\end{document}